\documentclass[usegraphicx,usenatbib,useAMS]{mn2e}
\newcommand\lsim{\mathrel{\rlap{\lower4pt\hbox{\hskip1pt$\sim$}}
        \raise1pt\hbox{$<$}}}
\newcommand\gsim{\mathrel{\rlap{\lower4pt\hbox{\hskip1pt$\sim$}}
        \raise1pt\hbox{$>$}}}

\newcommand{\enzo}{{\sc enzo}}
\newcommand{\Lya}{${\rm Ly\alpha}$} 

\usepackage{amsmath}
\usepackage{array}
\usepackage{appendix}

\usepackage{graphicx}
\usepackage{fixltx2e}
\usepackage{booktabs}
\usepackage{tikz,xcolor,hyperref}

\definecolor{lime}{HTML}{A6CE39}
\DeclareRobustCommand{\orcidicon}{%
  \begin{tikzpicture}
    \draw[lime, fill=lime] (0,0) 
    circle [radius=0.16] 
    node[white] {{\fontfamily{qag}\selectfont \tiny ID}};
    \draw[white, fill=white] (-0.0625,0.095) 
    circle [radius=0.007];
    \end{tikzpicture}
  \hspace{-2mm}
}

\foreach \x in {A, ..., Z}{%
  \expandafter\xdef\csname orcid\x\endcsname{\noexpand\href{https://orcid.org/\csname orcidauthor\x\endcsname}{\noexpand\orcidicon}}
}

\title[Detachment of ${H^-}$ by trapped \Lya]{Suppression of ${\bf H_2}$--cooling 
in protogalaxies aided by trapped {\bf \Lya} cooling radiation}
\author[J. Wolcott-Green et al.]
{Jemma Wolcott-Green$^{1}$\thanks{E-mail: jemma@astro.columbia.edu;
zoltan@astro.columbia.edu; gbryan@astro.columbia.edu}\orcidA, 
Zolt\'an Haiman$^{2}$\orcidB, and Greg L. Bryan$^{2}$\orcidC\\
$^{1}$Department of Physics, University of California Santa Barbara, MC 9530, Santa Barbara, CA 93106, USA\\
$^{2}$Department of Astronomy, Columbia University, 550 West 120th Street, MC 5246, New York, NY 10027, USA}

\begin{document}

\date{}

\pagerange{\pageref{firstpage}--\pageref{lastpage}} \pubyear{2012}

\maketitle

\label{firstpage}

\begin{abstract}
We study the thermal evolution of UV--irradiated atomic cooling haloes
using high--resolution three--dimensional hydrodynamic simulations.
We consider the effect of ${\rm H^-}$ photodetachment by \Lya~cooling
radiation trapped in the optically--thick cores of three such haloes, a
process which has not been included in previous simulations. ${\rm
  H^-}$ is a precursor of molecular hydrogen, and therefore, its
destruction can diminish the ${\rm H_2}$ abundance and cooling.
Using a simple high-end estimate for the trapped \Lya~energy density,
we find that ${\rm H^-}$ photodetachment by \Lya~decreases the
critical UV flux for suppressing ${\rm H_2}$--cooling by up to a
factor of $\approx 5$.  With a more conservative estimate of the
\Lya~energy density, we find the critical flux is decreased only by
$\sim 15-50$ percent.
Our results suggest that \Lya~radiation may have
an important effect on the thermal evolution of UV--irradiated haloes,
and therefore on the potential for massive black hole formation.
\end{abstract}


\section{Introduction}

It has long been known that ${\rm H_2}$ is the primary coolant 
in primordial gas at temperatures below a few thousand Kelvin 
\citep{Saslaw+Zipoy}. This has important implications for the 
first stars and protogalaxies, the reionization of the universe, 
and the formation of the first massive black holes \citep[see]
[for a review]{BYReview11}. Because ${\rm H_2}$ is easily 
photodissociated by soft-UV photons in the Lyman--Werner (LW) 
bands ($11.1-13.6$eV), radiation fields from the first stars can 
immediately have a strong feedback effect on their environments. 

Photodissociation of ${\rm H_2}$ has received particular
attention in the context of gravitational collapse of haloes 
with virial temperatures $T_{\rm vir} \gsim 10^4$K, in 
which gas is shock--heated to the virial temperature and 
can efficiently cool via atomic line cooling, even in the 
presence of a strong LW radiation field. These so--called 
atomic cooling haloes (``ACHs'') have been proposed as possible 
hosts of the first supermassive black hole seeds. A variety 
of studies have shown that the presence of a strong LW 
photodissociating flux can prevent ${\rm H_2}$--cooling 
during gravitational collapse in ACHs altogether, keeping 
the gas temperature near the virial temperature of the halo 
and thereby suppressing fragmentation on stellar--mass 
scales \citep[see][for a recent review]{IVH2020}.
Subsequent rapid accretion rates onto a dense core (${\rm M
\sim 0.1-1 M_\odot~yr^{-1}}$), enabled by the elevated gas 
temperature, may lead to the formation of a massive ($10^{4-6}
{\rm M_\odot}$) black hole seed via a supermassive star 
intermediary stage \citep{Hosokawa+12,Haemmerle+18}.

It is widely thought that in order for this so--called ``direct
collapse'' to occur, a large critical flux $J_{\rm crit}$ in the LW
bands is required to suppress ${\rm H_2}$--cooling \citep[but see
  also][]{Inayoshi+18,Wise+19}. Recent simulations have typically
found ${J_{\rm crit,21}=10^{3-4}}$, in the customary units $J_\nu =
J_{21} \times 10^{-21} {\rm erg~s^{-1}~cm^{-2}~ Hz^{-1}~sr^{1}}$ and
normalized at the Lyman limit
\citep{SBH10,Latif+14,Latif+15,Hartwig+15,Regan+14}. 
In general, the critical flux depends sensitively on the shape of 
the irradiating spectrum \citep{Sugimura+14,Agarwal+14,WGHB17},
${\rm H_2}$ self--shielding model \citep{WGHB11,Hartwig+15}, 
and rovibrational level populations \citep{WGH19}.

The effect of ``trapped'' Lyman--$\alpha$ (\Lya) photons on
fragmentation has also been considered in this context. Neutral
hydrogen column densities that build up during gravitational collapse
in ACHs exceed ${\rm N_H \gsim 10^{21}~cm^{-2}}$, and the resulting
large optical depth in the Lyman series lines can suppress atomic
cooling via \Lya~in particular. \citet{SS06} suggested that this could
lead to a stiffer equation of state than previously assumed, thus
suppressing fragmentation. However, subsequent studies have shown that
atomic cooling via other transitions, in particular H($2s \rightarrow
1s$), remain efficient even in dense cores where \Lya--cooling is
strongly suppressed \citep[e.g.][]{Schleicher+10}.

Recently, \citet[][hereafter JD17]{JD17}, used one--zone models 
to show that trapped \Lya~photons may instead alter the thermal 
evolution of collapsing ACHs via photodetachment of ${\rm H^-}$, 
an intermediary in the primary formation reaction for ${\rm H_2}$: 
\begin{equation}
{\rm H^- + H \rightarrow H_2 + e}. 
\end{equation}
${\rm H^-}$ can be destroyed by photons with ${\rm E > 0.76}$eV,
but previous studies have considered photodetachment only by 
the incident radiation field. JD17 found that while $Ly\alpha$
{\it photons alone do not suppress $H_2$ abundance enough to 
prevent molecular cooling}, when this additional photodetachment 
is included with an incident photodissociating flux, {\it the 
critical LW flux is decreased by up to a factor of $\approx5$.}
Such a reduction in $J_{\rm crit}$ would have important implications 
for the number density of direct collapse candidates, since the 
number of haloes exposed to a supercritical flux increases 
exponentially with decreasing $J_{\rm crit}$ \citep{Dijkstra+08}.

The goal of this study is to implement ${\rm H^-}$--photodetachment by
trapped \Lya~in a suite of three--dimensional hydrodynamic simulations
in order to further understand and quantify the magnitude of any
reduction of $J_{\rm crit}$ in atomic cooling haloes. We use the
publicly--available \enzo~code to simulate three such haloes with and
without the additional ${\rm H^-}$--photodetachment by trapped \Lya,
and further compare the results using different estimates for the
trapped \Lya~energy density produced during gravitational collapse.

In one set of simulations, we adopt the same approximation for the
\Lya~energy density as in JD17 and find a similar reduction in the 
critical flux as found in their one--zone models: 
$J_{\rm crit}$(\Lya)$/J_{\rm crit,0}\sim 0.2 - 0.8$. We also show, 
however, by post--processing the simulations, that their model may 
overestimate the amount of \Lya~produced in our haloes once the gas 
density reaches ${\rm n \approx 10-100~cm^{-3}}$, a key stage in the 
collapse determining whether ${\rm H_2}$--cooling is suppressed. 
Adopting a more conservative model for the trapped \Lya~intensity 
results in a more modest reduction in the critical flux, 
$J_{\rm crit}$(\Lya)$/J_{\rm crit,0} = 0.5-1$. Our results nevertheless suggest 
that trapped \Lya~may be important in the thermal evolution 
of UV--irradiated ACHs. A more detailed treatment of \Lya~radiative 
transfer is needed to precisely determine the photodetachment rate
and the resulting decrease in the critical flux.

This paper is organised as follows: We describe the numerical 
modeling in \S~\ref{sec:Model} and discuss the results in 
\S~\ref{sec:Results}. We summarize our primary findings and offer 
conclusions in \S~\ref{sec:Conclusions}.

\section{Numerical Modeling}
\label{sec:Model} 
\begin{figure}
  \includegraphics[height=3in,width=3.4in]{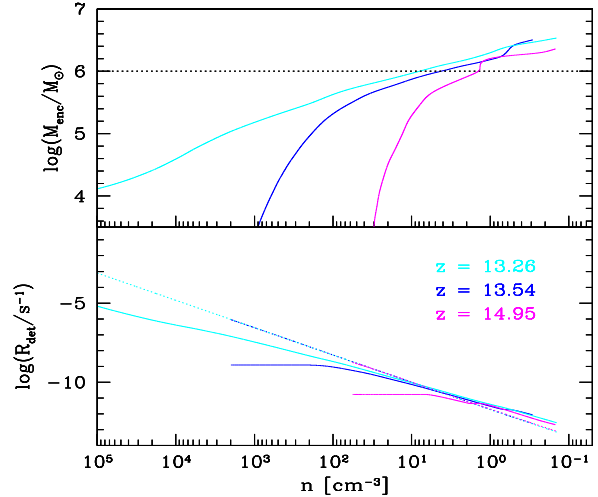}
  \caption{{\it Top:} Enclosed mass profile $M_{\rm enc}(<r,z)$ 
for Halo~B as a function of spherically averaged density 
$n(r)$, shown at $z_{\rm coll}$ (when the simulation 
reaches its maximum density), and two additional redshifts 
prior to collapse. {\it Bottom:} ${\rm H^-}$ photodetachment rate 
due to trapped \Lya~cooling radiation (see Equation \ref{eq:Rdet}) 
assuming a fixed mass, $M=10^6~{\rm M_\odot}$ (dotted), or
enclosed mass, $M_{\rm enc}(<r,z)$ (solid).} 
  \label{fig:Menclosed}
\end{figure}
\subsection{Numerical Modeling}
\label{subsec:Enzo}
We use \enzo\footnote{http://enzo-project.org}, a
publicly--available adaptive mesh refinement code, which 
uses an N-body adaptive particle mesh technique to follow
the dark matter (DM) dynamics, and a second-order accurate 
piecewise parabolic method to solve the hydrodynamics
\citep[see][for an in--depth description of the modeling]{Bryan+14}. 
We use the 9--species non--equilibrium chemistry network 
in \enzo~to follow the chemical evolution of gas with 
primordial composition. Radiative cooling by ${\rm H_2}$ 
is modeled with the cooling function from \citet{GP98}. We also 
updated several of the reaction rates in the default 
\enzo~chemistry network, as detailed in Appendix \ref{sec:Appendix}. 

Initial conditions for a simulation volume $1 h^{-1}$ Mpc 
on a side and $128^3$ root grid are generated with the 
{\sc music}\footnote{www-n.oca.eu/ohahn/MUSIC/} package
\citep{MUSICMethod11}. We initialize the simulation at $z_{\rm in}=99$
and adopt the cosmological parameters from the Planck 2018 
collaboration \citep{Planck18}, $\Omega_{\rm m} = 0.315$, 
$\Omega_\Lambda=0.685$ $\Omega_b= 0.0493$, $h= 0.674$, 
$\sigma_8 = 0.811$, and $n = 0.965$. 

In order to select haloes for ``zoom--in'' simulations, we run an 
initial low--resolution DM--only simulation from $z_{\rm in}=99$ to 
$z=10$, with a maximum of four levels of refinement. 
The {\sc rockstar} halo finder \citep{RockstarMethod13} is
run to find haloes with $T_{\rm vir} \gsim 10^4$K at $z=10$. 
Initial conditions are then re--generated with three nested 
grids enclosing the Lagrangian volume of the selected halo.
With the additional nested grids, the most--refined region 
has an effective grid resolution of $1024^3$ and dark matter 
particle mass $\sim 100~M_\odot$. 

High--resolution zoom simulations for three of the selected haloes 
are run from $z_{\rm in} = 99$ with the maximum refinement level 
set to 18, resulting in a minimum cell size of 0.0298 $h^{-1}$ cpc. 
The redshift when the simulation reaches this maximum refinement 
is referred to as the collapse redshift, $z_{\rm coll}$. 
In order to avoid numerical effects of discrete DM particles, the 
DM distribution is smoothed at a maximum refinement level of 13. 
Cells are flagged for additional spatial refinement when the baryon 
or dark matter mass is four times greater than that of the most 
refined cell. In addition, the local Jeans length is always resolved 
by at least 16 cells in order to avoid spurious fragmentation \citep{Truelove+97}. 
The properties of all three haloes at their collapse redshift with 
$J_{21} > J_{\rm crit}$ are shown in Table \ref{tbl:HaloProperties}.

%
\begin{table}
\begin{center}
\caption{Mass and virial temperature of Haloes~A-C at the collapse 
redshift with $J=J_{\rm crit}$ (no ${\rm H^-}$ photodetachment by \Lya.)}
\label{tbl:HaloProperties}
\begin{tabular*}{0.45\textwidth}{@{\extracolsep{\fill}}l l l l}
\hline\hline
 & Halo A & Halo~B & Halo~C \\
  \hline
  $z_{\rm coll}(J_{\rm crit}$) & 11.87 & 13.26 & 9.36 \\
  $M_{\rm tot}/10^7 {\rm M_\odot}~(z_{\rm coll})$  & 1.9 & 1.4 & 2.1  \\ 
  $T_{\rm vir}/10^3$K ($z_{\rm coll}$) & 7.9 & 7.2 & 6.7  \\ 
  \hline\hline\\
\end{tabular*}
\end{center}
\end{table}

\subsection{Implementing ${\bf H^-}$ photodetachment by \Lya}
\label{subsec:Lya}
In our first set of simulations including ${\rm H^-}$ photodetachment 
by trapped \Lya, we utilize the model described by \citet[][see their 
Equations 3--7]{JD17}, and briefly summarized here. They assume the energy 
radiated in \Lya~cooling radiation balances the gravitational binding 
energy released by a cloud of mass $M = 10^6~{\rm M_\odot}$ collapsing 
on a free-fall timescale. 

The derived \Lya~energy density $u_\alpha$ accounts for the increased
path length of a photon escaping an optically--thick medium,
\begin{equation} 
u_\alpha = M_F \times \frac{L_{\rm Lya} r_{\rm cloud}}{V_{\rm cloud} c},
\end{equation}
where
$L_{\rm Lya}$ is the luminosity from the simple toy model above,
$r_{\rm cloud}$ is the size of the cloud,
$V_{\rm cloud}$ is the geometrical volume of the cloud,
$c$ is the speed of light, and
$M_F\sim a_v\tau_{\rm Ly\alpha}$, is the dimensionless path length boost. 
For this boost, $a_v$ denotes the Voigt profile,
 $\tau_{\rm Ly\alpha} = 5.9 \times 10^6(\frac{N_{\rm H}}{10^{20}~{\rm cm}^{-2}})
  (\frac{T}{10^4 {\rm K}})^{-\frac{1}{2}}$
is the line-center optical depth, and  $N_{\rm H}$ is the neutral hydrogen column density,
found by assuming a cloud of uniform density.
For an isotropic 
\Lya~field within the cloud and using the cross--section for photodetachment, 
$\sigma_{\rm H^-} = 5.9 \times 10^{-18}~{\rm cm^2}$ at ${\rm E_{\rm Ly\alpha} 
= 10.2eV}$, they derive the photodetachment rate:
\begin{multline}
R_{\rm det} \simeq 10^{-8} {\rm s}^{-1}
\left( \frac{M}{10^6~{\rm M_\odot}} \right) ^{10/9}
\left( \frac{T}{10^4~{\rm K}} \right) ^ {-1/3}\\
\times \left( \frac{n}{10^2~{\rm cm^{-3}}} \right) ^{31/18}
\left( \frac{\it B_\alpha}{2} \right).
\label{eq:Rdet}
\end{multline}
Here $n$ is the density, and the parameter $B_\alpha$ is included to capture the
possible impact of density gradients and
non--uniform
diffusion of spatial diffusion of \Lya~photons.
These gradients and non-uniform diffusion could increase the trapped
Ly$\alpha$ energy density in the center (see Appendix in JD17),
but for our purposes, it is treated as a free parameter. The set of
simulations we run with this rate (directly from JD17) will be
referred to as ``constant mass'' models.

In order to evaluate the validity of this one--zone model prescription
for our simulated haloes, we have examined the enclosed mass profiles
at several redshift snapshots in our haloes up to the collapse
redshift.  These are shown for one of the haloes in the top panel of
Figure \ref{fig:Menclosed}. Because the density profiles toward the
core are relatively steep, the enclosed mass $M(<r,z)$ at $n \gsim
10^2 ~{\rm cm^{-3}}$ falls rapidly below $10^6~{\rm M_\odot}$. As a
result, the photodetachment rate with fixed ${\rm M=10^6~M_\odot}$, as
in JD17, is significantly larger than if the actual enclosed mass
$M(<r,z)$ is instead used in Equation \ref{eq:Rdet}, as shown in the
lower panel of Figure \ref{fig:Menclosed}.

Using $R_{\rm det}$ with $M(<r,z)$ self--consistently in the
simulation would unfortunately require that we compute the density
profile on the fly, which is beyond the scope of the current
modeling. Instead, we run a set of simulations assuming $M =
M(<r,z_{\rm coll})$, which we will refer to as the ``enclosed mass''
models. While this is still larger than $M(<r,z)$ at earlier
redshifts, it is sufficient for the purposes of the relatively simple
model used to estimate the \Lya~radiation field.

\subsection{Incident Radiation Field}
\label{subsec:LW}
\begin{figure}
  \includegraphics[height=3in,width=3.4in]{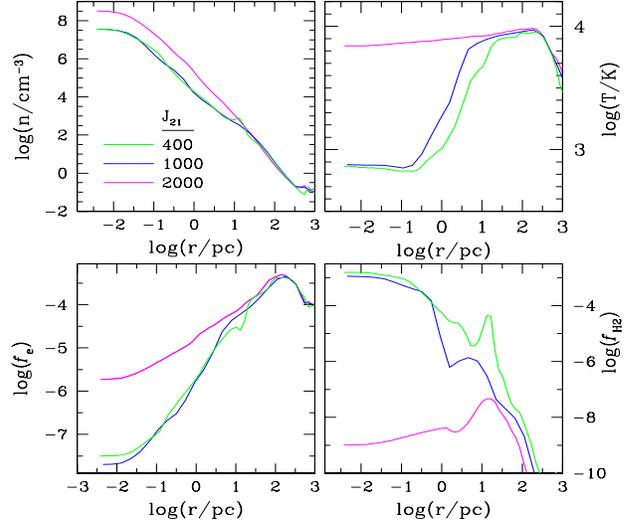}
  \caption{The profiles of spherically--averaged density (upper 
left), temperature (upper right), electron and ${\rm H_2}$ 
fractions (lower left and right, respectively) for Halo A. 
All profiles are at the collapse redshift of each simulation, 
and for varied intensity of the incident Lyman--Werner radiation: 
$J_{21}= 400,1000,3000$.} 
  \label{Fig:Profiles}
\end{figure}
\subsubsection{Photodissociation of Molecular Hydrogen} 
We adopt the commonly--used approach for modeling an incident 
${\rm H_2}$--photodissociating flux with a blackbody 
spectrum with ${\rm T_* = 10^5}$K up to the Lyman limit.
Ionizing photons are assumed to have been 
absorbed\footnote{We do not include the characteristic 
saw--tooth modulation seen in the cosmological LW background 
spectrum as a result of absorption in the IGM \citep{HAR00}. 
The critical LW flux, $J_{\rm crit}$, is much larger than 
the expected cosmological background, and is most likely to 
originate instead from a bright near neighbor galaxy 
\citep{Visbal+14,Regan+17}.}, likely by neutral gas within 
the irradiating galaxy itself. While the Pop III IMF remains 
uncertain, the ${\rm H_2}$ photodissociation rate derived 
with this spectrum is a good approximation for metal--free 
starburst populations \citep[e.g.][]{WGHB17}. 
We use the fitting formula for the optically--thick ${\rm H_2}$ 
photodissociation rate from \citet{WGHB11} in order to directly
compare to the JD17 results; note that this fit was recently
updated by \citet{WGH19} to significantly improve the accuracy 
for vibrationally warm ${\rm H_2}$ (${\rm T\gsim 3000}$K, 
$n\gsim 10^3~{\rm cm^{-3}}$).
The self--shielding ${\rm H_2}$ 
column density is estimated with a local ``Sobolev-like'' length 
as the characteristic length scale: 
\begin{equation}
L_{\rm char} = \frac{\rho}{\nabla \rho}, \\ 
\end{equation}
\begin{equation}
{\rm N_{H_2} = n_{H_2}} \times L_{\rm char}. 
\end{equation}
This has been previously implemented in the \enzo~network and 
\citet{WGHB11} showed it is a more accurate local prescription 
than the oft--used Jeans length. 

\subsubsection{Photodetachment of $H^-$ by Incident Radiation Field} 
While ${\rm H^-}$ photodetachment by the incident flux is not 
the dominant mechanism for ${\rm H_2}$--suppression with a
$10^5$K blackbody spectrum, it is included in our modeling 
with the standard rate coefficient: $k_{\rm H^-} = 1.07 \times 
10^{-11} J_{21}~{\rm cm^3~s^{-1}}$.

\section{Results and Discussion}
\label{sec:Results}
\begin{table}
\begin{center}
\caption{Critical fluxes in units $J_{\rm crit,21}/10^3$ for Haloes~A-C, 
with and without ${\rm H^-}$ photodetachment by trapped \Lya. Results 
are shown for constant mass, $M=10^6 {\rm M_\odot}$, and enclosed mass, 
$M(<r,z_{\rm coll})$, models.
The top row shows the value of the factor $B_\alpha$, which scales
the trapped Ly$\alpha$ photon density to allow for non-uniform
density profiles and photon-diffusion (see Eq.~\ref{eq:Rdet}).}
\label{tbl:Jcrits}
\begin{tabular*}{0.45\textwidth}{@{\extracolsep{\fill}}l l l l l l l}
  \hline\hline
\end{tabular*} 
\begin{tabular*}{0.43\textwidth}{@{\extracolsep{\fill}}l l l l}
  &          & constant mass & enclosed mass \\
  \cline{3-4}
\end{tabular*}
\begin{tabular*}{0.45\textwidth}{@{\extracolsep{\fill}}l l l l l l l}
$B_\alpha$& 0         &   1  &    10  &    1    &   10    \\
  \hline\hline
Halo A  &   6         &   3   &    2   &   5    &   3    \\
Halo~B  &   12        &   4   &    2   &   11   &   10   \\
Halo~C  &   7         &   6   &    2   &   7    &   4    \\
  \hline\hline\\
\end{tabular*}
\end{center}
\end{table}

\begin{figure*}
  \centering
  \label{fig:PhaseDiagrams}
  \begin{tabular}{cc}
    \includegraphics[width=65mm]{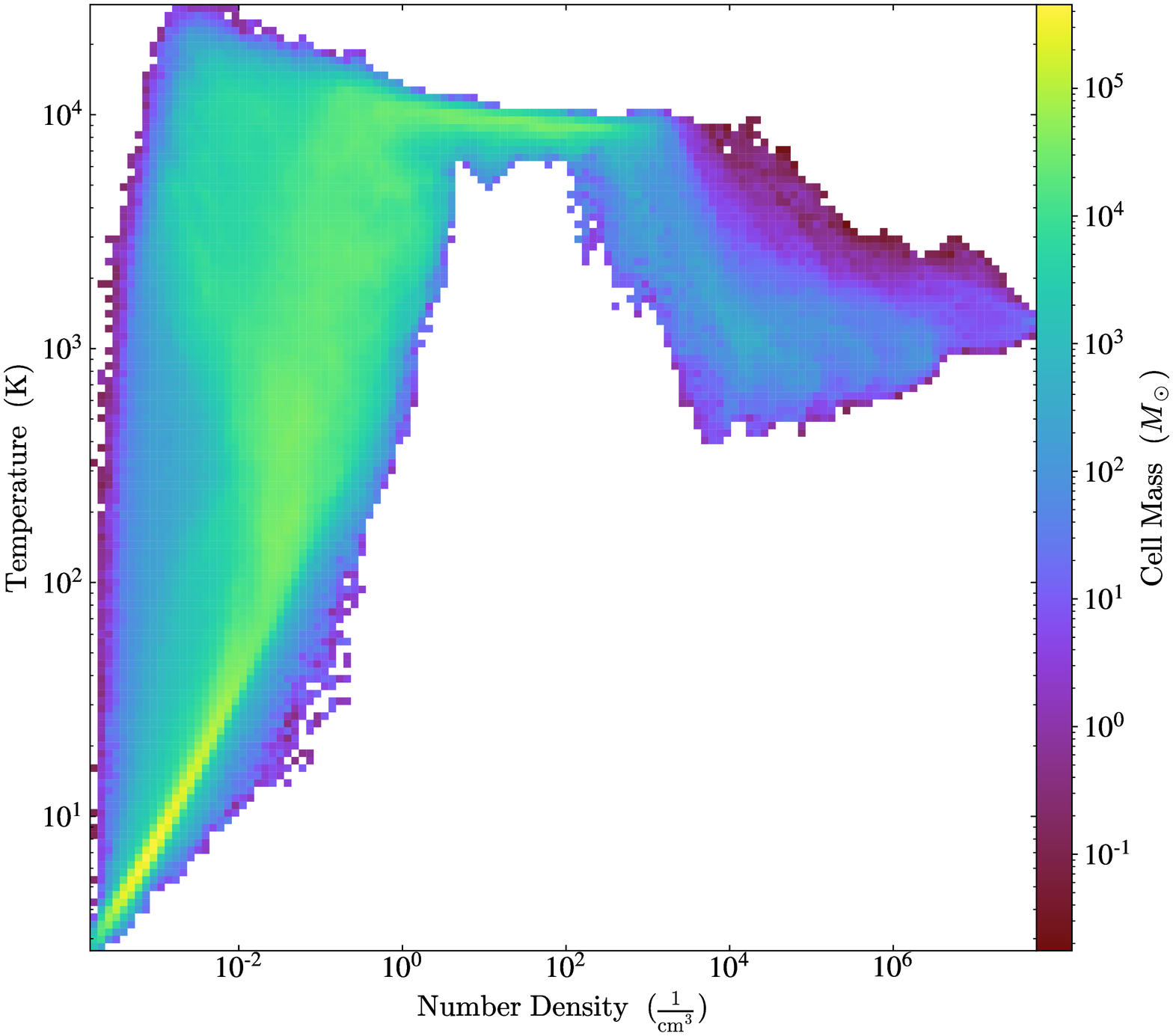}&
    \includegraphics[width=65mm]{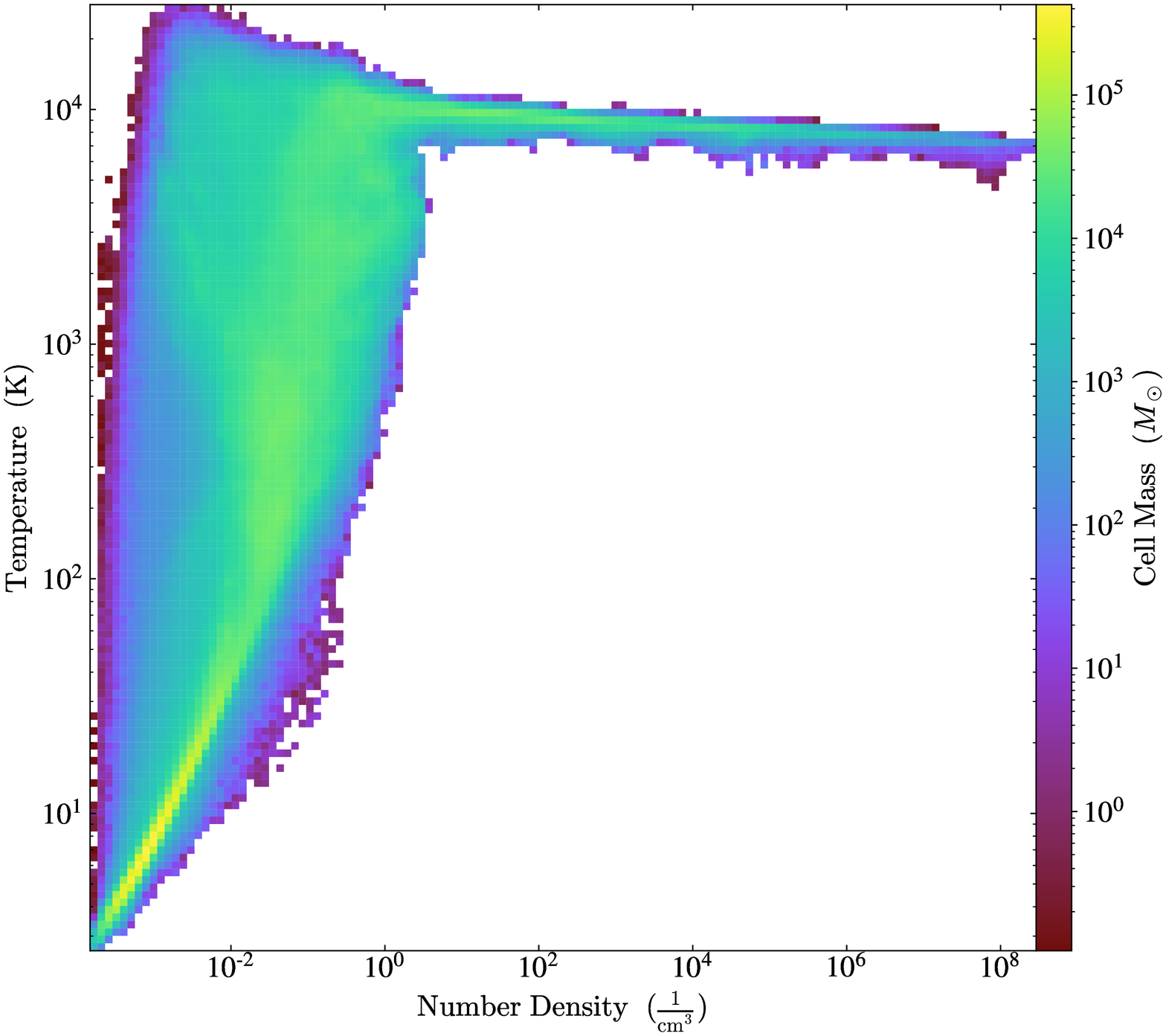}\\
  \end{tabular}
\caption{Phase diagrams of the sub--critical (left) and super--critical 
(right) runs in Halo~A shown at the collapse redshift ($B_\alpha=1$).}
\end{figure*}

\subsection{Impact of ${\rm H^-}$ Detachment by Trapped Ly$\alpha$}

In order to determine the critical flux for each of the haloes 
and \Lya~models, we run the zoom simulations for each with a 
series of incident flux strengths. The initial runs with 
$J_{21} = 10^3, 5\times 10^3, 10^4$ were analyzed at the 
collapse redshift to determine if ${\rm H_2}$--cooling was 
suppressed. Subsequently, a set of more finely--spaced flux 
tests (increments of $10^3$ in $J_{21}$) were run to precisely 
determine $J_{\rm crit}$ required to prevent cooling below 
$T\approx T_{\rm vir}$. 

The resulting critical fluxes for each of our haloes and \Lya~
models are listed in Table \ref{tbl:Jcrits}. In the $B_\alpha=0$ 
cases, the only ${\rm H^-}$--photodetachment is from the incident 
radiation. For these, $J_{\rm crit,21}$ in the three haloes is 
found to be in the range (6-12)$\times 10^3$. This is a factor of
$\sim 5-10$ larger than the one--zone results in JD17. Previous
studies which have also found a larger critical flux in simulations 
as a result of hydrodynamic effects including shocks, which can 
increase the ionization fraction, and are not captured by the 
one--zone modeling \citep[e.g.][]{SBH10,Latif+14,Latif+15}. The 
halo--to--halo $J_{\rm crit}$ variation is also consistent with 
previous studies, which is often found to be within a factor of 
$\sim$ three. 

We show in Figure \ref{Fig:Profiles} the spherically--averaged
density and temperature profiles\footnote{We use the publicly--available
package $yt$ \citep{ytMethod11} for simulation data analysis 
and visualization; see yt--project.org.} at the collapse redshift for 
one of our simulated haloes, as well as the fractional abundances of 
electrons and ${\rm H_2}$. Each panel shows the results (in Halo~A) 
for varied $J_{21} = (0.2, 0.5, 1.0) J_{\rm crit}$($B_\alpha=10)$. 
The results follow the typical pattern seen in previous simulations 
of LW--irradiated ACHs: with sub--critical flux, the ${\rm H_2}$--fraction 
in the dense core reaches the standard ``freeze--out'' value $\sim 10^{-3}$ 
\citep{OH02} resulting in robust ${\rm H_2}$--cooling and gas  
temperatures of a few hundred Kelvin in the inner $r \sim 0.1$pc. 
Once the critical flux is reached, the ${\rm H_2}$--fraction 
is suppressed, $f_{\rm H_2}\lsim 10^{-7}$, and the gas temperature 
remains near the virial temperature of the halo ${\approx 7000}$K.

\subsection{Constant Mass Models} 
In the ``constant mass'' models, the photodetachment rate by trapped 
\Lya~is identical to that implemented by JD17 (Equation \ref{eq:Rdet} 
above, $M = 10^6~{\rm M_\odot}$); for direct comparison to JD17, we
run two sets of models with  $B_\alpha=1$ and 10. The critical flux 
in the $B_\alpha=1$ case is decreased by a factor of 2-3 in Haloes~A 
and B, while in Halo~C it is reduced by only $\sim 15$ percent 
compared to $B_\alpha=0$. The latter is similar to the $\sim 18$ per 
cent reduction found by JD17 in their one--zone models. In our models 
with $B_\alpha = 10$, the critical flux is decreased further: 
$J_{\rm crit}(B_\alpha=10)/J_{\rm crit,0}=0.33,0.17,0.29$, in Haloes~A, 
B, and C, respectively. This is as expected, since the \Lya~detachment 
rate is larger, and is also consistent with the JD17 results, in 
which $J_{\rm crit}(B_\alpha=10) \approx 0.18 J_{\rm crit,0}$. 

Example phase diagrams of number density and temperature at 
$z_{\rm coll}$ (for Halo~A and $B_\alpha=1$) are shown in 
Figure~\ref{fig:PhaseDiagrams}; in the left panel the flux was 
sub--critical ($J_{21}=2/3~J_{\rm crit}$) and the right panel 
shows results with a super--critical flux. These too are 
consistent with the results of previous studies; in particular, 
the sub--critical case shows that the shock--heated gas remains 
at $\sim T_{\rm vir}$  during the collapse until the density 
reaches $10^{2-3}~{\rm cm^{-3}}$, at which point the ${\rm H_2}$ formation 
time becomes smaller than the dissociation time--scale, and the 
gas then cools \citep[see, e.g.,][for an in--depth discussion of 
the relevant timescales determining $J_{\rm crit}$]{SBH10}.

\subsection{Enclosed Mass Models} 
The decrease in $J_{\rm crit}$ is smaller in our ``enclosed mass'' 
models, for which the \Lya~energy density is calculated with 
$M(<r,z_{\rm coll})$ (derived from post--processing the haloes
run with ${\rm B_\alpha=0}$). This is as expected, since the 
enclosed mass in the region where $n \gsim 10~{\rm cm^{-3}}$ is 
less than ${\rm 10^6 M_\odot}$ (see Fig \ref{fig:Menclosed} and 
Table \ref{tbl:HaloProperties}); 
therefore, this model yields a smaller \Lya~energy density than
the constant mass case at the stages of collapse that are key 
for determining the critical flux. Here, with $B_\alpha = 1$, 
$J_{\rm crit}$ is decreased in only two of the haloes (A and B) 
and very modestly (by $\sim 10-15$ per cent). In the $B_\alpha=10$ 
case, all three haloes see a decrease in $J_{\rm crit}$, ranging 
from $\sim 15-50$ per cent (see Table \ref{tbl:Jcrits}).

\begin{figure}
  \includegraphics[height=3in,width=3.4in]{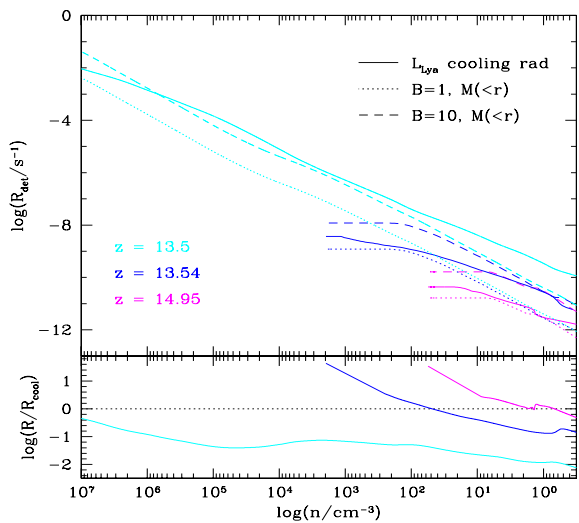}
  \caption{{\it Top:} The photodetachment rate of ${\rm H^-}$ for
    Halo~B at several redshifts, using Equation~\ref{eq:Rdet} with the
    enclosed mass and $B_\alpha = 1,10$ (dotted and dashed lines,
    respectively). The solid line shows the same using the volume
    integrated cooling rate to derive a \Lya~ luminosity (rather than
    based on gravitational binding energy release in a free-fall time,
    as in JD17). {\it Bottom:} ratio of the photodetachment rate using
    the $B_\alpha=1$, $M(<r,z_{\rm coll})$ model (dotted cyan line in
    the upper panel) to that from the \Lya~cooling radiation
    calculated at each redshift (solid lines in the upper panel). This
    comparison is shown because the former is used in one set of
    simulations (see Table \ref{tbl:Jcrits}).}
  \label{fig:Rcooling}
\end{figure}

Even with the modified rate using the enclosed mass, this is a
somewhat crude model for estimating the \Lya~energy density.  For the
sake of a ``sanity check,'' in Figure~\ref{fig:Rcooling} we show the
photodetachment rate with \Lya~energy density obtained directly from
the volume--integrated atomic cooling rate (rather than based on the
gravitational binding energy released in a free-fall time, as
previously). As shown in the top panel, this ``$R_{\rm cool}$'' rate
(solid lines) mostly lies between our enclosed-mass models with
$B_\alpha=1$ and 10 (dotted and dashed lines, respectively). The
exception is at the final snapshot, $z_{\rm coll}$, when $R_{\rm
  cool}$ is significantly larger than even the $B_\alpha=10$ model at
densities below $\sim 100~{\rm cm^{-3}}$.

This rough agreement with the $B_\alpha=1,10$ rates is reassuring 
that the model employed here yields a reasonable estimate for 
$R_{\rm det}$; however, as discussed in \S~\ref{subsec:Lya}, the 
rate implemented in our simulations actually is obtained using 
$M(<r,z_{\rm coll})$, since we do not track the enclosed mass on 
the fly (see \S~\ref{sec:Model}). Therefore, the rates implemented 
in our \enzo~network are the z$=13.5$ curves (cyan) with $B_\alpha=1$ 
and 10 (dotted and dashed). 

The ratio of the $B_\alpha=1$ rate with $R_{\rm cool}$ is shown in the
lower panel of Figure \ref{fig:Rcooling}. At the pre--collapse
redshifts, where $n_{\rm max} \approx 10^{2-3}~{\rm cm^{-3}}$, our
implemented rate with $M(<r,z_{\rm coll})$ diverges from $R_{\rm
  cool}$ at $n \sim 10-100 ~{\rm cm^{-3}}$ and becomes $\sim$ an order
of magnitude larger at the highest densities.  By the time the
collapse has reached $n_{\rm max}= 10^{7}~{\rm cm^{-3}}$ at $z_{\rm
  coll}$, $R_{\rm cool}$ is much smaller than the original rate with
$B_\alpha=1$. This, suggests that our models may underestimate the
trapped \Lya~intensity, especially at very high densities.  A more
detailed study of the radiative transfer is needed in order to more
precisely determine the photodetachment rates in a collapsing halo.

\subsection{Gas Inflow Rate and Mass of the Final Object}
\label{sec:Mdot}
The rate of gas inflow onto the core in ACHs is a key factor 
in determining the mass of the central object that can form 
\citep[e.g.][and citations therein]{IVH2020}. A ``critical'' 
mass inflow rate required for SMS formation has been found 
to be $\sim 0.05 {\rm M_\odot~{\rm yr}^{-1}}$ 
\citep{Hosokawa+13,Schleicher+13,Haemmerle+18}. 

\begin{figure}
  \includegraphics[clip=true,trim=0.8in 0.5in 0in 0in,
height=3in,width=3.4in]{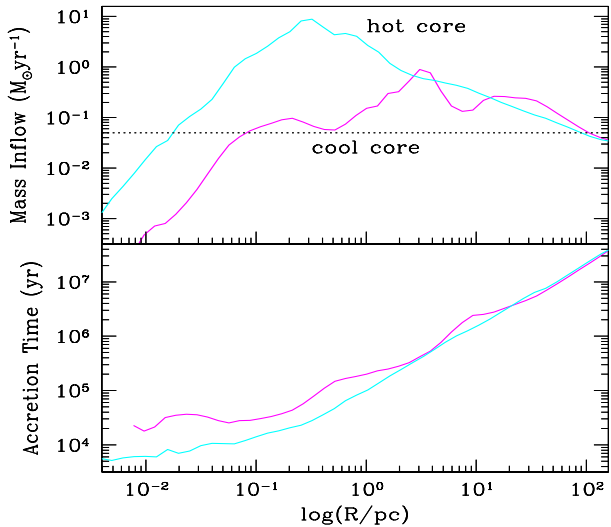}
  \caption{
{\it Lower panel:} The local accretion time-scale 
${\rm \equiv R/\langle v_{rad} \rangle}$is shown for Halo A 
at the collapse redshift ($z=11.9$) in the case 
of a supercritical flux, which prevents ${\rm H_2}$--cooling
(cyan curve), and subcritical flux (magenta curve) in which
the gas is able to cool via ${\rm H_2}$. 
{\it Upper panel:} Rate of mass inflow 
(${\rm \dot M = 4\pi R^2 \rho} \langle v_{\rm rad} \rangle$) is 
shown for the same halo snapshot as the lower panel. The ``critical'' 
inflow rate is marked with a dotted horizontal line. At 
the relevant radii for direct collapse, $\sim 0.1-0.3$pc 
(see \S~\ref{sec:Mdot}), the inflow safely exceeds this threshold
rate only in the case of the supercritical flux (hot core).
The results are qualitatively similar for the other haloes
and therefore have been omitted here for clarity.}
  \label{fig:Mdot}
\end{figure}

In Figure \ref{fig:Mdot} we show the mass inflow rate (upper panel)
(${\rm \dot M = 4\pi R^2 \rho} \langle v_{\rm rad} \rangle$) for Halo A 
at the collapse redshift in the case of a supercritical (cyan) and 
subcritical (magenta) flux. As expected, in the supercritical 
case (with $T_{\rm gas} \sim T_{\rm vir}$) the mass inflow rate
is significantly higher (by up to two orders of magnitude) 
than in the subcritical case, in which the gas has cooled via
${\rm H_2}$. 

In the lower panel of Figure \ref{fig:Mdot}, the local accretion
time--scale ($t{\rm _{acc} \equiv R/\langle v_{rad} \rangle}$) 
is shown for the same halo snapshot. For metal-free gas 
contracting on a Kelvin--Helmholtz time-scale of $\sim 10^{4-5}$ 
years, the relevant radii, where ${t_{\rm acc} \gsim t_{\rm KH}}$, 
are $\sim 0.1-0.3$pc. At these radii, the mass inflow rate safely 
exceeds the critical rate only in the case of the supercritical 
flux. These results are consistent with previous studies, which 
typically find that haloes in which ${\rm H_2}$--cooling is 
suppressed are more likely to maintain high accretion rates and 
accumulate up to ${\rm 10^{4-5}~{\rm M_\odot}}$ of gas within 
the Kelvin--Helmholtz time.

\subsection{Depletion of Ly$\alpha$ by vibrationally warm $H_2$}

There are several ${\rm H_2}$ Lyman transitions that lie close to the
\Lya~line center; therefore, in a gas with a significant ${\rm H_2}$
fraction, \Lya~can be systematically converted to ${\rm
  H_2}$ fluorescent emission.  \citet{Neufeld90} showed that a large
fraction of \Lya~photons are thus converted when the ${\rm v=2}$,
${\rm J= 5,6}$ states are thermally populated and the gas temperature
is $\gsim$ a few thousand Kelvin.  For example, in a cloud with ${\rm
  N_H=10^{20}~cm^{-2}}$ and ${\rm H_2}$ fraction $f_{\rm H2}=10^{-3}$,
$> 90$ per cent of \Lya~(emitted by a central source) are converted to
${\rm H_2}$ Lyman band radiation (via the B-X 1-2P[5] and B-X 1-2R[6]
transitions), before they can escape the cloud.

JD17 assume that this process is unimportant since the ${\rm H_2}$ 
fraction in gas exposed to a near--critical flux is small, 
$f_{\rm H2}\sim 10^{-7}$. Further, they point out that even if 
\Lya~photons are absorbed by vibrationally--warm ${\rm H_2}$, most 
of these events will result in a fluorescent radiative cascade, 
releasing additional photons that can photodetach ${\rm H^-}$. We
note that it is also possible that this \Lya~pumping of ${\rm H_2}$
could directly contribute to the photodissociation rate and thus 
further suppress the ${\rm H_2}$ abundance; however, a detailed 
accounting of this process is beyond the scope of this work.

\section{Conclusions}
\label{sec:Conclusions}

We have run a suite of high--resolution 3D hydrodynamic simulations to
study the effect of trapped \Lya~cooling radiation on the thermal
evolution of UV--irradiated atomic cooling haloes.  We show that the
critical UV flux for suppressing ${\rm H_2}$--cooling is decreased by
up to a factor of $\approx 5$ when ${\rm H^-}$ photodetachment by 
\Lya~is included with a simple high-end estimate of the trapped 
\Lya~photon density.
In models with a more conservative estimate of the trapped
\Lya~energy density, we find the critical flux is decreased by $\sim
15-50$ per cent.  Our results are consistent with previous one--zone
models \citep{JD17} and suggest that \Lya~radiation may have an
important effect on the thermal evolution of UV--irradiated haloes.
While we have implemented two different models for the trapped 
\Lya~energy density, there remains significant uncertainty due to 
the difficulty of accurately computing this quantity on--the--fly 
in simulations. This should be addressed in future work through a 
more detailed treatment of \Lya~radiative transfer. 

\section*{Acknowledgments}
This material is based upon work supported by the National Science 
Foundation under Award No. 1903935. JWG is grateful to Cameron Hummels 
for useful discussions about the simulation set--up.
ZH and GB acknowledge support from NSF grant NNX15AB19G.
GLB acknowledges support from NSF grants AST-1615955 and OAC-1835509.
\bibliography{paper}
\clearpage                                                                                                                                
\appendix
\section{Updated Chemistry Rates}
\label{sec:Appendix}

Our chemistry model includes the following updates to
the standard \enzo~network.\\

\noindent {\it Collisional dissociation of ${\it H_2}$ by H:} \\
We utilize the \citet{MSM96} fit for collisional dissociation 
of ${\rm H_2}$, 
\begin{equation}
{\rm H_2 + H \rightarrow H + H + H},
\end{equation}
including the contribution from dissociative tunneling,
which has not previously been used in the \enzo~network.
\citet{Glover15a} notes this term becomes larger than direct 
dissociation at temperatures below $4500$K, and found that 
neglecting it leads to $J_{\rm crit}$ determinations that 
are erroneously large by a factor of $\sim$ two. \\ 

\noindent {\it Associative Detachment of ${\it H^-}$ with H:} \\
We use the updated rate coefficient from \citet{Kreckel+10} 
for the associative detachment reaction: 
\begin{equation}
{\rm H^- + H \rightarrow H_2 + e^-}.
\end{equation}
In the sensitivity study by \citet{Glover15b}, this is among 
the five most important reactions determining $J_{\rm crit}$. 
The rate from \citet{Kreckel+10} is in good agreement with other
recent determinations; however, \citet{Glover15b} found that 
the $25$ per cent systematic uncertainty results in $\sim 40$ 
per cent uncertainty in $J_{\rm crit}$.  \\

\noindent{\it Radiative recombination of $H^+$}: \\ 
We use the Case B rate from Hui+Gnedin'97 for the radiative 
recombination reaction, 
\begin{equation}
{\rm H^+ + e^- \rightarrow H + \gamma}.
\end{equation}
This has been is previously included as an option in \enzo, 
but not always used in primordial chemistry models 
\citep[e.g.][]{Abel+97,SBH10}. In the context of an atomic 
cooling halo, where the mean-free path for ionizing photons 
is generally small, using the Case B rate is appropriate. 
\citet{Glover15b} note that the Case A rate is $\sim 60$
per cent larger in the relevant temperature range and 
therefore causes $J_{\rm crit}$ to be $\sim 80-90$ per cent
smaller in their models.\\

\noindent{\it Radiative association of H and $e^-$}:\\ 
We replaced the \citet{Hutchins76} rate previously used in the 
\enzo~network for radiative association reaction: 

\begin{equation}
{\rm H + e^- \rightarrow H^- + \gamma}.
\end{equation}

We instead use the \citet{Abel+97} rate agrees well with 
alternative analytic fits from \citet{Stancil+98,GP98} 
in the range ${\rm T =  10^2 - 10^4}$K, while the 
\citet{Hutchins76} rate is significantly larger than the 
other three at ${\rm T \gsim 3000}$K. \citet{Glover15b} 
find that $J_{\rm crit}$ results using the \citet{Hutchins76}  
rate are nearly a factor of two larger than in models 
using any of the other three rates. \citet{Glover15b}
also note that the \citet{Hutchins76} is not valid in
the conditions of interest for determining $J_{\rm crit}$ 
in ACHs, ${\rm n \sim 10^3~cm^{-3}}$, $T \sim 7500$K, and 
therefore recommends against using it in this context. 

\label{lastpage}

\end{document}